\documentclass[twocolumn]{IEEEtran}
\usepackage{booktabs}
\usepackage{authblk}
\usepackage{url}
\usepackage{silence}
\usepackage[font=scriptsize]{caption}

\usepackage[pdftex]{graphicx}
\usepackage{epstopdf}
\usepackage{pifont}
\usepackage[table,xcdraw]{xcolor}
\usepackage{array}
\usepackage{verse}
\usepackage{longtable}
\usepackage{supertabular}
\usepackage{cite}
\usepackage{color}
\usepackage{amsmath}
\usepackage{lettrine}
\usepackage{hyperref}
\usepackage{multirow}
\usepackage{dirtytalk}

\hypersetup{
    colorlinks=true,
    linkcolor=blue,
    filecolor=magenta,
    urlcolor=cyan,
}

\begin{document}

\title{How 5G (and concomitant technologies) will revolutionize healthcare}
%\author{Siddique Latif, Junaid Qadir, Shahzad Farooq}
\author[1,2]{Siddique Latif}
\author[1]{Junaid Qadir}
\author[3]{Shahzad Farooq}
\author[4]{Muhammad Ali Imran}
\affil[1]{Information Technology University (ITU), Punjab, Lahore, Pakistan}
\affil[2]{National University of Sciences and Technology (NUST), Islamabad, Pakistan}
\affil[3]{Nokia Networks, Helsinki, Finland}
\affil[4]{School of Engineering, University of Glasgow, United Kingdom}
\maketitle

\begin{abstract}

 In this paper, we build the case that 5G and concomitant emerging technologies (such as IoT, big data, artificial intelligence, and machine learning) will transform global healthcare systems in the near future. Our optimism around 5G-enabled healthcare stems from a  confluence of significant technical pushes that are already at play: apart from the availability of high-throughput low-latency wireless connectivity, other significant factors include the democratization of computing through cloud computing; the democratization of AI and cognitive computing (e.g., IBM Watson); and the commoditization of data through crowdsourcing and digital exhaust. These technologies together can finally crack a dysfunctional healthcare system that has largely been impervious to technological innovations. We highlight the persistent deficiencies of the current healthcare system, and then demonstrate how the 5G-enabled healthcare revolution can fix these deficiencies. We also highlight open technical research challenges, and potential pitfalls, that may hinder the development of such a 5G-enabled health revolution.

\end{abstract}


\section{Introduction} 

%``Health is among the most important conditions of human life and a critically significant constituent of human capabilities which we have reason to value.''--Amartya Sen.

Good health has a constructive effect on all aspects of human and social well-being including personal happiness, workforce productivity, and economic growth.  Recognizing the importance of healthcare, facilitating affordable universal access to healthcare is already enshrined as an important goal of the United Nations' new Sustainable Development Goals (SDG) that defines the UN's development agenda for the next $15$ years. In the words of the Nobel Laureate Amartya Sen, ``\textit{Health is a critically significant constituent of human capabilities which we have reason to value}''. It has been shown in literature that investment in healthcare pays huge dividends. In the Economists' Declaration, originally launched in $2015$ with $267$ high-profile economist signatories, world-leading economists called on global policymakers to plead for a pro-poor pathway to universal health coverage as an essential pillar of sustainable development\footnote{\url{http://universalhealthcoverageday.org/economists-declaration/}}. A case was made that healthcare investments make perfect economic sense since according to the Global Health $2035$ report by the Lancet Commission on Investing in Health, every dollar invested in the healthcare of poor countries has a nine-fold or higher return. 

%Sound health is one of the most significant constituent of human capabilities that critically influences all aspects of human life. It forges the level of human capital and education by reducing incapacity and debility. Avoiding any health incidents is particularly critical in poor developing countries because poor health can compound and exacerbate poverty since one unhealthy episode in a family can spin it into a trap that the family can never get out of. 

\begin{figure}[!ht]
\centering
\captionsetup{justification=centering}
\centerline{\includegraphics[width=.5\textwidth]{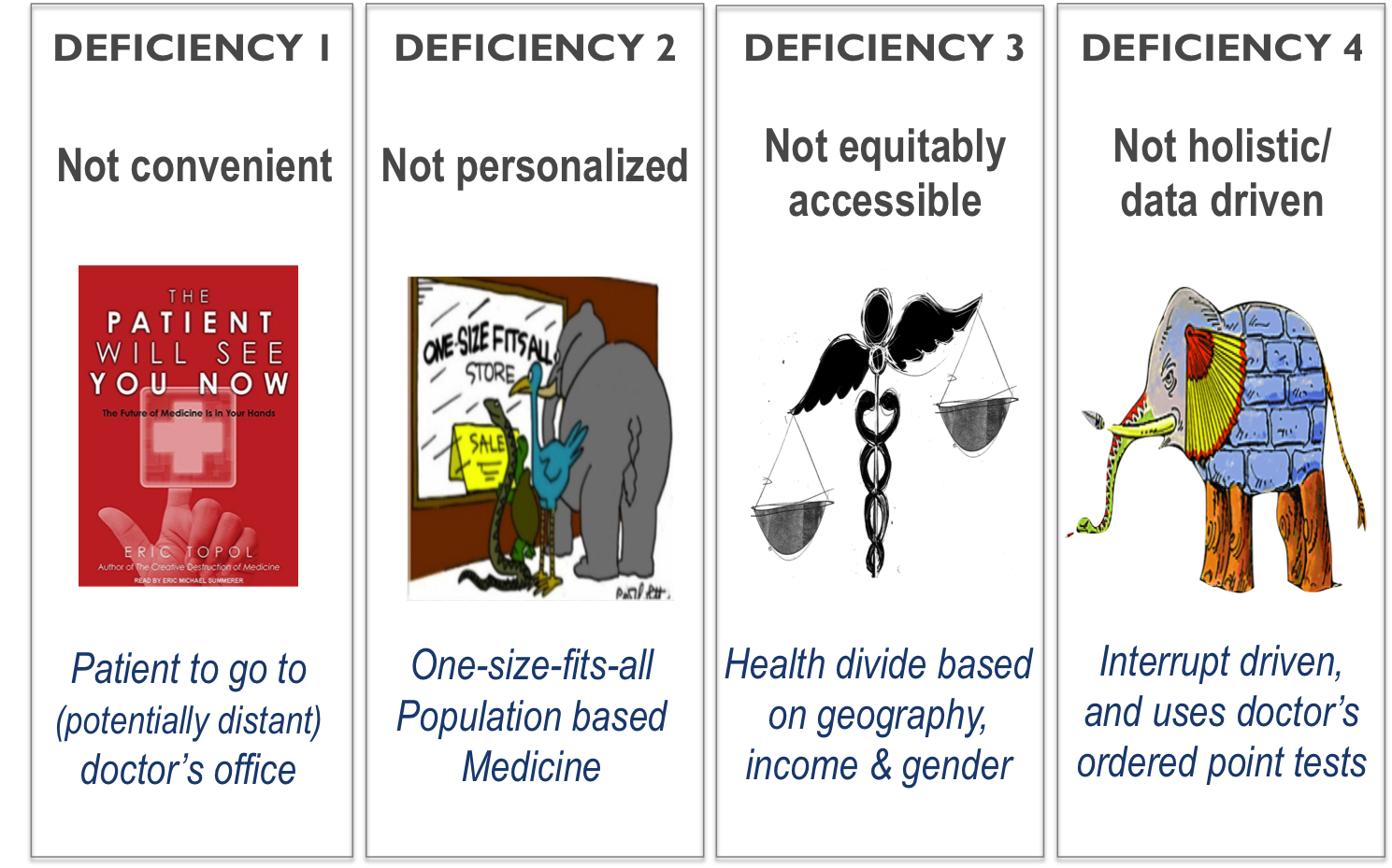}}
\caption{The four major deficiencies of conventional healthcare systems}
\label{fig:4-fold}
\end{figure}

Despite the core role of human health in human development and progress, today's healthcare system is largely dysfunctional and in the need of a major overhaul. Broadly speaking, the ills of the healthcare system can be categorized into four major deficiencies (illustrated in Figure \ref{fig:4-fold}).

\textit{Firstly}, the current healthcare system is \textit{not convenient} for patients since the current healthcare system is not patient-centric. As an example, the patient has to go, or be taken, to a doctor's office or a hospital for any non-trivial illness which is inconvenient for the patient (who would likely prefer to rest) and also for the patient's caregivers (e.g., the patient's guardian or family member who must take the patient to the clinic). The patients also need to slot in their health related appointments in their busy schedule and this sometimes lead to carelessness in giving due attention to regular and required health-checks with doctors. 

\textit{Secondly}, the current healthcare system is \textit{not personalized} according to the individual patient. Doctors prescribe medications based on population averages rather than the individual characteristics. As today, it is very difficult and costly to adopt tailored treatments based on individuals' medical history and genetic profile.   

\textit{Thirdly}, the current healthcare system is \textit{not equitably accessible}. Similar kinds of healthcare facilities are not equally accessible to patients or utilized by only a certain groups of people, based on their ethnicity, socioeconomic status, and geographic residence, etc. Similarly, lack or limited access to basic healthcare services are putting patients (especially disabled people) at much higher risk and causing adverse health outcomes. 

\textit{Fourthly}, the current healthcare system is \textit{not holistic/data-driven}.
The Institute of Medicine, a division of the National Academy of Sciences, representing our most prestigious scientists and physicians, published the report To Err Is Human, which proclaimed that ``at least 44,000 people, and perhaps as many as 98,000 people, die in hospitals each year as a result of medical errors that could have been prevented,” but which arose because ``faulty systems, processes, and condition'' led people either to make mistakes or to fail to prevent them. Beyond the human toll, these errors cost between $\$17$ billion and $\$29$ billion.

Although modern medicine is ripe with numerous success stories (such as the eradication of diseases as smallpox, invention of antibiotics and anesthesia, development of modern surgery and therapy techniques), the overall healthcare industry has been largely impervious to a technological revolution. This has been the case due to many reasons such as its highly regulated and policy-driven nature as well as the unique nature of its value chain unlike other markets, in which someone makes choices (e.g., a doctor), someone else is a consumer and user (e.g., patient), and someone else altogether pays (e.g., the insurer or the government via taxpayer). 
%Around the globe, healthcare challenges are in various forms and shapes, putting tremendous pressure on current systems. The healthcare value-chain is critically unique in which someone chooses (e.g. the doctor), someone uses (e.g. the patient) and someone pays (e.g. the insurer). Such a scheme is rarely found in other markets.
The Institute of Medicine (IoM), a division of the US National Academy of Sciences (NAS), summarized the ills of the US healthcare system as follows: ``\textit{If banking were like health care, automated teller machine (ATM) transactions would take not seconds but perhaps days or longer as a result of unavailable or misplaced records. If home building were like health care, carpenters, electricians, and plumbers each would work with different blueprints, with very little coordination. If shopping were like health care, product prices would not be posted, and the price charged would vary widely within the same store, depending on the source of payment. If automobile manufacturing were like health care, warranties for cars that require manufacturers to pay for defects would not exist. As a result, few factories would seek to monitor and improve production line performance and product quality. If airline travel were like health care, each pilot would be free to design his or her own preflight safety check, or not to perform one at all}'' \cite{mcginnis2013best}.

While the point of the charge sheet above is certainly not that healthcare should function precisely in the way that other industries do, indeed each industry is different from others. However, the point is that the banking, construction, retailing, automobile manufacturing, flight safety, public utilities, and personal services have developed certain best practices in terms of quality assurance, accountability, and transparency that healthcare industry should also incorporate. 

More than anything else, the healthcare industry needs to be reoriented so that the patient becomes the core concern of the system. In such a patient-centric healthcare, patient will be empowered with information and the preferences and convenience of patients will automatically be incorporated. Furthermore, payment incentives should be redesigned so that outcomes and values are rewarded and not a volume of procedures. Finally, with a transparent system, errors will be promptly identified and corrected and a data-driven approach will become routine allowing for continuous improvement through reflections on past and evidence based healthcare decisions.

In this paper, we show how 5G and various other technologies such as IoT, big data analytics, artificial intelligence (AI) and machine learning (ML) will restructure healthcare system. We believe that 5G network will address not only personal communications but also aims to create a fully digital society. In which sensors can be embedded in tissue (pacemaker), ingestible (smart pills), epidermal (smart skin or digital tattoo), wearable (clothing or jewelry), and external (traditional blood pressure monitors and smart watches). ML algorithms will estimate appropriate micro-dosages of insulin to be delivered by the pump, as well as to detect anomalies that might be forwarded to human experts who can ensure that no medical problem has occurred. Telemedicine or e-Health will enable resource pooling by remote consultation and remote surgery. Surgeons will have haptic feedback (robot touches) from the patient tissues. Similarly, patients will able to measure their own vitals at a fraction of cost and with great convenience. People in rural and low-income zones will have equitable health at reduced cost. In this way, overall healthcare will be transformed. In addition, this paper also discusses various destructive innovation in healthcare and technology challenges and pitfall that will come up with 5G.

The paper is organized as follows. In Section \ref{De}, we discuss the various challenges posed by the current healthcare system. In Section \ref{fix}, we highlight how technology can fix the existing lacks of the healthcare system and disruptive innovations are also presented. In section \ref{5G} opportunities of 5G for healthcare are discussed in details, followed by technology challenges and pitfall for healthcare in section \ref{tech}.  Finally, we conclude this paper in Section \ref{CON}.

%What will 5G be? \cite{andrews2014will}
%Fundamentals of LTE \cite{ghosh2010fundamentals}
%5G: Personal mobile internet beyond what cellular did to telephony %\cite{fettweis20145g}
%5G for Personalized Health and Ambient Assisted Living %\cite{hermens20165g}

%M-Health Solutions Using 5G Networks and M2M Communications %\cite{de2016m}

\section{Challenges Posed By The Current Healthcare System}
\label{De}

The current healthcare system is stressed by a number of challenges including an aging population; the rising disease burden of lifestyle-related chronic diseases (such as diabetes, high blood pressure, etc.); the absence of patient-centered scalable clinical operating models; the lack of healthcare facilities and human resources (or limited access thereto);  and the high costs associated with the provision of high-quality care \cite{challenges}. Some of these and other major global healthcare challenges are discussed in more detail next.

%Therefore, long-term thinking and strategic investments are required that pays attention not only to allocating funds for improving healthcare, but also on public health policies, and on the interactions of the health sector with every other sector. It also aims for higher medical research, changing clinical practices, improving human capital in healthcare and most importantly on the utilization of innovative technologies in healthcare system. 

\subsection{Challenges with EHRs}
Electronic Health Record (EHR) is a repository containing patients' digital data that is stored and exchangeable securely to multiple authorized users. It stores retrospective, prospective, and concurrent health information with the purpose to support efficient, continuing, and quality service in integrated health \cite{hayrinen2008definition}. In contrast to the virtues and success of EHRs, there are various challenges and limitations. The lack of interoperability is a major problem because hospitals and physicians are mostly not connected. This causes the patients' health information to be constrained within hospitals and laboratories. 

\subsection{Lack of Universal Access}
Universal access to healthcare entails that everyone can have equitable access to health services without any discrimination especially on the capacity to pay \cite{agarwal2012universal}. Coverage of healthcare services is limited when a country lacks in trained health care professionals, services, and equipment; available resources are not located in proximity; and individuals are unable to afford services due to their high cost \cite{savedoff2009moving}. Universal healthcare services is particularly challenging for developing and underdeveloped countries, where health resources and practitioners are in short supply, particularly, in rural areas  \cite{latif2017mobile}. Universal access can be achieved by progressively eliminating the above-mentioned challenges that prevent people from having fair and comprehensive health facilities determined at the national or international level. % A functioning healthcare system is fundamentally important to achieve the universal coverage of healthcare around the globe. %The current healthcare system failed to attain global coverage because of various deficiencies due to which people suffer and unnecessarily die. For example, each year $98000$ people die in hospitals because of preventable medical errors which arose by faulty systems, processes, and conditions \cite{topol2013creative}. 

\subsection{The Long-Term Chronic Care Burden}

Chronic diseases are increasing globally and have become the most dominant and serious threat. By $2020$, $157$ million people in the United States are estimated to have at least one chronic illness and the cost will increase to $80$\% of total healthcare expenditures ($75$\% in $2000$) \cite{wu2000projection}. The current global healthcare system is particularly troubling for people with chronic illness \cite{priester2005current}. People suffering from chronic diseases rely more heavily on the healthcare system: they utilize the system more often, consume extra resources, visit multiple doctors, and have long-term relationships with them. Therefore, when the healthcare system fails, patients with chronic disease are more affected. 

\subsection{Challenges for Aging Populations}
The world's population is rapidly growing older, leading us to a higher number of elderly people in our society. The number of people aged $65$ or older is projected to increase from $8$\% ($524$ million) of world's population in $2010$ to $16$\% ($1.5$ billion) in $2050$ \cite{world2011global}. The increasing share of the elderly population with increased life expectancy is changing the cause of death from infectious diseases to chronic noncommunicable illnesses. These demographic shifts are posing enormous challenges to healthcare systems. In near future, the current healthcare systems will fail to provide long-term care to older people with multiple chronic illnesses. Changes are required to make strategies for older adults to live independently by providing
high-quality care.

\subsection{Resources Constraints}

Despite the great success in creating impressive outlook of healthcare system, the overall healthcare services are still insufficient. %In the global context, many general issues related to available healthcare resources are evident and imminent. %The scarcity of resources with the focus to produce more with less is a ever-present issues for healthcare organizations. The most striking factors are the and distribution of the healthcare facilities and available workforce. 
Patients have to travel from distant place to visit their doctors. Traveling barriers cause rescheduling or missing of appointments, and delay in medications. There are $57$ countries with a critical shortage of healthcare workers, for instance, Africa has $2.3$ health workers per $1000$ population as compared to Americas, which have $24.8$ health workers per $1000$ population \cite{naicker2009shortage}. Similarly, in Pakistan, there is one doctor for $1038$ inhabitants \cite{latif2017mobile}. The problem is now becoming more acute and the world will have an anticipated shortage of $12.9$ million health-care workers by $2035$ \cite{world2014global}. Therefore, we need some serious developments in healthcare to increase the productivity of health professionals by using telemedicine and e-Heath-like services.

%In addition to the shortage of health workers  These consequences may cause poor chronic disease management that results in poor health outcomes.

%In addition, disease like cancer is expected to grow significantly due to the aging of population. This will cause a substantial increase in economic burden of cancer. For example, if cancer incidence, survival, and required costs for care remained constant by 2020, the costs of cancer care in USA is projected to increase from \$124.5 billion dollars in 2010 to \$157.8 billion dollars in 2010 \cite{mariotto2011projections}.

\subsection{Problems with Healthcare Information Systems}

The advancements in information communications technology have potentials to bring a significant transformation in the healthcare system by connecting medical devices, automating financial transactions and preventing errors to enhance consumer confidence in the health system. %There are ranges of areas in healthcare system which can be improved with the employment of information technologies, but health industry is slow to invest and embrace such innovative technologies \cite{greiner2003challenges}. %Therefore, healthcare system has not been revolutionized to the same degree as technologies have changed other aspect of society.  
But the healthcare systems are very complex, as they include communication and processing of heterogeneous health information, optimal allocation of available resources and their administrative management simultaneously. Existing wireless technologies (3G, 4G, and WiMAX) exploit macro-cells to provide wider range suitable for lower data rates. These technologies are mainly suitable for smart health monitoring devices, social interactions, and wellness monitoring applications \cite{istepanian2016m}. For healthcare systems, we require a heterogeneous wireless technology with multiple frequencies ranges that can provide guaranteed high data rate and very low latency for medical services like remote surgery. % diagnostic with using large medical data (i.e., medical images and videos). 

\subsection{Lack of Data Driven Culture}

In current healthcare paradigm, patients are assessed on population averages and data-poor practices, and it is practically not possible to conduct standard parallel group randomized controlled trials (RCTs). Moreover, clinical evidence generated by non-standard RCTs has poor generalizability, therefore have limited applicability to patients, and even also restricted to compare the effectiveness of drugs and medical devices \cite{brook2009possible}. %Doctors prescribe medications based on population averages rather than individual characteristics. Decisions are made based on departmental norms. Patients are assessed for the appropriateness of medications. 
%In addition, on average safe and effective treatments may deliver uneven benefits, risks,  and can cause adverse effects to individual patients (called heterogeneity in treatment effects) \cite{kent2010assessing}. 
In such data-poor system, the medication doses are often over- or underestimated---that can result in adverse drug reactions. This represents a crucial missed opportunity to embrace the development of an intelligent and evidence-based healthcare system that generate and apply the best evidence-based care to each patient; discover the natural outgrowth of patient care; and also ensure innovation, safety, quality, and add values to the healthcare system.

%Today, doctors prescribe medications based on population averages rather then individual characteristics. Decisions are made based on departmental norms. Patients are assessed for the appropriateness of medications. In such data-poor system, the medication doses are often over- or underestimated---that unforeseen the adverse drug reactions. Similarly, physicians also ask for various lab tests before diagnosing a patient that is problematic for patients in critical situation. In addition, available health information is unstructured, exists in medical prescription notes, biometric data, computerized physician order entry (CPOE), clinical registries, electronic patient records (EPRs), and insurance claims,  etc. The healthcare organizations are not in the position or have the capability to process this massive volume of data in a timely fashion to improve health outcomes \cite{mathew2015big}. In order to provide data analytic based solutions, cultural changes are  required within a healthcare system, that enables an enterprise-wide, progressive, sustainable and automated environment. 

 %Therefore,  we need such automated platform that aims to improve health at the population level and makes population health management feasible, sustainable, and scalable. 

\subsection{Healthcare Disparities}

Currently, healthcare systems are mostly income-based instead of need-based. People who need healthcare services crucially are getting less access than the people who need it least. For example, on average in United State rich are the biggest buyers of healthcare services despite being healthier than the poor \cite{dickman2016health}. In the other study, it is found that people in the lower incomes quartile have poorer health \cite{wolfe2011poverty}. This dramatic shift in healthcare is creating disparities in health outcomes across income groups.

\section{How technologies can fix the current healthcare system}
\label{fix}

 In this section, we will discuss how technologies such as Internet of Things (IoT), big data for healthcare, (wireless connectivitiy), and other disruptive health innovations can fix the current broken healthcare system.

\subsection{Various health advances with IoT}

The Internet of Things (IoT) is an abstraction of infinite, smart, physical, and virtual objects that have unique identities connected to create an ultimate cyber-physical pervasive framework. These devices capture, monitor and transmit data to public or private cloud to facilitate a new level of convenient and efficient automation. We are witnessing the rise of cellular Internet of Things (IoT) and it is expected that there will be $20$ billions of devices or things connected to the Internet by $2020$ \cite{bahga2014internet}. IoT trend is a next-generation technology that can change the whole business spectrum with a variety of applications such as smart cities, industrial control, retails, waste management, emergency services, security, logistics.  Most importantly, IoT is considered as an enchanting technology that can revolutionize the current healthcare system with a variety of cutting-edge and highly individualized solutions such as remote health monitoring, remote diagnostics, tele-auscultation, chronic diseases management, independent care for elderly and much more \cite{islam2015internet}. Patients compliance with medication and treatment by healthcare providers is another prominent potential application of IoT. In addition, IoT can also be used to authenticate medicine, monitor drugs supplies and efficient scheduling of available resources to ensure their best use for more patients.

In medical IoT, various medical sensors, devices, smart phones, imaging devices, PDAs and EHRs act as a core part of the system. These devices monitor important health information, like physiological vital signs, changes in mood and behaviors, blood glucose, that can be effectively utilized by healthcare providers to improve the quality of care and health outcomes. Furthermore, IoT-based solutions have potentials to reduce the required time for remote health provision, increase the quality of care by reducing costs with enriched user's experience. %The most important factors hindering the growth of IoT market include lack of clarity in health data communication standards and protocols, privacy and security concerns, and regulatory issues. 

%\begin{figure}[!ht]
 %\centering
  %\captionsetup{justification=centering}
  %\centerline{\includegraphics[width=.5\textwidth]{IOT.eps}}
  %\caption{IoT healthcare services and applications}
  %\label{fig:IOT}
%\end{figure}

%Figure. \ref{fig:IOT} shows various kind of services and applications of IoT that aim to improve healthcare system in terms of productivity and outcomes by providing automated and connected care. The service (in Figure. \ref{fig:IOT}) is a generic term that has potential to play role in a set of solutions and applications. For example, a service \say{clinical data gathering} can be used for various applications such as child health monitoring, health information charting, and remote monitoring, etc. 

%Internet of Medical Things for cardiac monitoring: Paving the way to 5G mobile networks \cite{jusak2016internet}
%How 5G technology enables the health internet of things \cite{west20165g}

\subsection{Big Data for Healthcare}

 In the last few years, we are increasingly living in a digital world where devices like smart phones, EHRs, biomedical and wearable sensors produce a large volume of health data. Such data can be referred to as ``big data'' due to its high-velocity and wide variety holds a lot of promise for evidence-based human developmental efforts \cite{ali2016big}. The digitization of human being and the rise of big health data which allows the remote and continuous monitoring of each heartbeat, moment-to-moment blood pressure, oxygen concentration in blood, body temperature, glucose, human activities. and emotions. Human health data and behavior information can also be used for analytics to gain deep insights into the various aspects of human life. For example, when the information about human activities, their geographical location, shopping habits, travel patterns, and social circle are used together with health information such as health records and genetic information, it enables the discovery of latent population-based health patterns and the efficicacy of different intervention methods.

  %\begin{figure}[!ht]
 %%\captionsetup{justification=centering}
  %\centerline{\includegraphics[width=.35\textwidth]{BigData.eps}}
  %\caption{Patient Centric Big Data enabled Healthcare Ecosystem}
  %\label{fig:BDC}
%\end{figure}

Even though big data analytics has now changed almost every sector in the global economy, it has not efficiently contributed in healthcare \cite{gulamhussen2013big}. Researchers and stakeholders from the healthcare industry are struggling to exploit the unprecedented predictive power of big data in healthcare system \cite{costa2014big}. The most pressing challenge is the lack of patients' trust to share their personal data, as patients are not in control of their health information. To overcome this barrier, patients must be put in control over their personal information by providing them full visibility and control in big data ecosystem. The new strategies must be formed to stimulate active, legal and performance based models. This will encourage open information sharing and vast variety of opportunities for healthcare system.

% Such systems can be designed by exploiting the effectiveness of Big Data that can handle and correlate a heterogeneous, continuous form of behavioral and health data for millions of people and also provide a wide variety of services and applications that will make drug therapies more effective, and enable detection of new drug interactions more quickly. We enlist various services and applications of Big Data analytic in healthcare ecosystem in Figure. \ref{fig:BD}. 

%\begin{figure}[ht]
 %\centering
  %%\captionsetup{justification=centering}
  %\centerline{\includegraphics[width=.5\textwidth]{BigDataApp.eps}}
  %\caption{Big Data in healthcare system}
  %\label{fig:BD}
%\end{figure}

%6 V's of Big Health Data
%Exhaust data (CDR) applications
%Big data in biomedicine \cite{costa2014big}

%Big data: The next frontier for innovation, competition, and productivity \cite{manyika2011big}
%Revolutionizing medicine and Public Health \cite{pentland2013revolutionizing}
%From big data analysis to personalized medicine for all: challenges and opportunities \cite{Alyass2015}
%Health-CPS: Healthcare cyber-physical system assisted by cloud and big data \cite{zhang2015health}

\subsection{(Wireless) Connectivity}
The advancements in wearable computing, bio-engineering, mobile devices have enabled a dramatic increase in the exploitation of ubiquitous and pervasive wireless technologies in the healthcare system. The use of wireless technology for healthcare has been explored in various studies \cite{lazakidou2010wireless, khoumbati2009handbook}. For example, Ng et al.  \cite{ng2006wireless} present how wireless technologies and medical services can be integrated to provide flexible, convenient and economical telemedicine. Originally, telemedicine is a growing application of wireless technologies that enable the electronic exchange of data, voice, text, images and other information to provide expert-based medical care from a distance.% (see Figure. \ref{fig:tele}) \cite{watfa2011healthcare}.
%\begin{figure}[ht]
 %\centering
  %\captionsetup{justification=centering}
  %\centerline{\includegraphics[width=.4\textwidth]{Tele}}
  %\caption{A general architecture of telemedicine with two locations}
  %\label{fig:tele}
%\end{figure}

Nowadays, the rapid growth of wireless sensor networks, wireless communications, wearable sensors and especially smart phones have paved the path to new e-Health systems, superior to telemedicine. Future technology like Ultra-Reliable Low latency Communications (URLLC) brought by 5G  will make health communications more resilient and will open many new healthcare opportunities like remote surgery and remote diagnosis with haptic feedback. Moreover, simultaneous wireless information and power transfer (SWIPT) can result in a significant gain in the performance of medical implants.   Another promising application of 5G is bio-connectivity which will decentralize hospitals services and enable the provisioning of medical care on the move (i.e., for emergency response in ambulances) \cite{mahmoodi20175g}. This will open a new era of advantages from enhancing performance in hospitals to new ways of monitoring the patients' health, disease progression, and individualized pharmaceutical analysis. We will discuss more details of 5G-related healthcare and applications in Section. \ref{5G}.

%In order to make e-Health a reality, an enduring cooperation is needed, among several research areas of engineering including bioengineering, wireless sensors, Body Area Network (BAN), data mining, microprocessors, and wireless as well as wired communications. %The future e-Health systems come with some new challenges related to security and privacy of health records; designing of small devices with power constraints; digitization of existing medical information; enabling interoperability of data;  and the network with low latency, high availability and low packet loss.    

%\begin{figure}[ht]
 %\centering
  %\captionsetup{justification=centering}
  %\centerline{\includegraphics[width=.45\textwidth]{Wireless.eps}}
  %\caption{Wireless technologies and their applications in healthcare}
  %\label{fig:W}
%\end{figure}

%Some applications:
%\begin{itemize}
 %   \item Telemedicine (healing at a distance)
%Some applications of telemedicine: (1) Low-cost video conferencing; (2) %Tele-radiology; (3) Tele-ECG
%Tele-dermatology; (4) Tele-pathology; (5) Tele-psychiatry
 %   \item Tactile Internet \cite{fettweis2014tactile}
  %  \item Remote Surgery
%\end{itemize}

\subsection{Artificial Intelligence/Machine Learning/Probabilistic modeling}	

Traditional healthcare systems often give suboptimal health outcomes despite the best efforts of the clinicians and physicians. The incomplete knowledge of patients' health, and the personal biases of clinicians, account for most of these suboptimal health outcomes. Optimal results require such systems that are driven by scientifically and  statistically validated data to provide benefit choices to the population. In the near future, it is inevitable that the majority of classical or traditional physicians' practices, prescription methods, and monitoring systems techniques will be replaced by intelligent devices, software toolbox, and testing procedures. Healthcare will mature into more consistent and more scientific in terms of delivering a better quality of care with the expense of AI and data gathering techniques. This will provide continual monitoring, more rigorous, personalized, precise as well as logical care. This evolution from an exclusively human-based system to the intelligent and automated expert systems will provide a bionic assist to doctors by substantially complementing or enhancing their expertise. This will make even the average nurse or physician to be performed at the level of best specialists \cite{khosla201420}. 

As an example of the advances in ML and AI, we point out the 
significant advancements the ML technique of \textit{deep learning} has made, particularly in image processing and recognition. There is now growing hope that deep learning and other ML models will help improve healthcare treatment and diagnostic procedures by recognizing patterns that are too subtle for the human eye. Deep learning is already being used to detect skin cancer in images with almost the same number of errors as made by professional dermatologists\footnote{\url{https://www.technologyreview.com/s/513696/deep-learning/}}. Healthcare systems based on AI, machine learning, and probabilistic models have potentials to provide therapeutic recommendations, prognosis learning, personalized real time risk scoring, etc. There is now growing interest, among doctors and entrepreneurs to largely deploy deep learning techniques.

\subsection{Disruptive Health Innovations}

Technological innovations---e.g., implantable devices, point-of-care (POC) testing, robotic surgery\footnote{\url{http://uchealth.com/services/robotic-surgery}}, and 3D printing\footnote{\url{https://www.3dsystems.com/industries/healthcare}}---are changing the landscape of current healthcare systems in terms of diagnostics, treatments, and delivery of quality care. Advances in big data, ubiquitous computing, semiconductors, and nanotechnology are creating profound opportunities for disaster recovery coordination; epidemics prediction, evaluation of new medicine, and large-scale DNA sequencing to detect human genetic variation. Some of these important technologies are discussed next. 

%\subsubsection{Genetics}

%SOPHiA Genetics\footnote{http://www.sophiagenetics.com/} is mostly used data-driven medicine based on advanced artificial intelligence. It uses DNA sequencing and mathematics to unlock the secrets of most difficult diseases, such as Cystic Fibrosis, different types of cancer and various more conditions. SOPHiA Genetics works with more than 300 leading hospitals and provides excellent help to diagnoses of Oncology, Hereditary cancer, cardiology, metabolism, and pediatrics. 

\subsubsection{Ingestible Sensors}

The ingestible sensors allow physicians to monitor patient's ingestion and adherence of medicines in real time. These sensors are activated after the ingestion and transmit information about drugs ingestion and its relation with patients' behaviors such as physical activity, and physiological responses (blood pressure, heart rate, and sleep quality). They are being used to monitor medication adherence \cite{hafezi2015ingestible}, like adherence to tuberculosis therapy \cite{belknap2013feasibility} and highly reliable monitoring of timing and intake of drugs for clinical management of kidney transplant patients. 

\subsubsection{Wearable Sensors to Wireless Charging Implants}

Recent advancements in sensing technologies have made great progress in monitoring patients' health in an unobtrusive manner. %that is transforming episodic, and manual sampling processes to context aware and continuous sensing. 
%These sensing technologies are playing an important role in achieving the prime focuses of the future healthcare system which include prevention from diseases as well as early detection and invasive management of illnesses. 
The wearable devices are widely being used to measure various physiological signs and physical activities. Recent advances in microelectronics and nanofabrication have shifted the trend from wearable sensors to implants---that transmit health related information from inside the body. These implantable sensors have enhanced capabilities to capture vital health related signs along with critically accelerated detection of failing implants thereby minimizing healthcare hazard \cite{andreu2015wearable}. SWIPT has the potential to ease sterilization, reduce procedure and maintenance time by increasing reliability for these implants. Now, it is not too distant future when a patient having severe memory loss due to injury or Alzheimer's stroke will be able to create long-term memories with the help from electronic implant\footnote{\url{https://www.technologyreview.com/s/513681/memory-implants/}}. 

\subsubsection{Robot Assisted Therapy and Surgery}

Medical robots and computer-aided surgery have great potentials to fundamentally change the nature of medicine and surgery. They work as the patient-specific information driven surgical tools that empower the surgeons to treat the patients with great safety, reduced morbidity, and improved efficiency. 
The most striking efforts by robot-assisted surgery can be seen in stereotactic brain surgery, surgery in gynaecology, microsurgery, endoscopic surgery, and orthopaedics surgery as well as for providing assistance to nurses in labor intensive tasks\cite{taylor2008medical,liu2014robot,ding2014giving}. Robotic prosthetics is also revolutionizing healthcare by creating neurally controlled artificial organs (i.e., limb) that can restore near-natural motor and sensory capabilities of amputee patients.

%The information like medical imaging (computed tomography (CT), X-ray, Ultrasound, positron emission tomography (PET), magnetic resonance imaging (MRI), video etc.), lab tests and other information about human anatomy physiology and disease specification are used to generate a powerful computer representation of the patient. The fundamental capability of robots to transform complex information into the physical useful task is being utilized to perform surgical tasks and also for augmentation in human capabilities.% Moreover, robotic technologies are also being utilized in providing assistance to nurses in labor intensive tasks (lifting and repositioning patients), thereby preventing them from musculoskeletal injuries \cite{ding2014giving}. 
     %Robotic nursing assistants deliver core benefits associated with safe patient lifting, reduce patient falls and injuries because of improper lifting. 
%They have an expert learning system that senses the center of gravity, while lifting the patients, and also able to adjust its position. Therefore, patients feel more secured as compared to a lifting and repositioning involving human hands. 

\subsubsection{Open Source EHR/EMR Systems}

The open source EHR or electronic medical record (EMR) is very crucial to improve the efficiency of present resource-constrained healthcare systems.% by creating global healthcare community. %with the help of scalable, robust, user-driven, open source platform.
Open source software is gaining attention for their adoption in healthcare industry that is helping to overcome several barriers (i.e., excessive cost, lack of interoperability, and the transience of vendors). The fully integrated and secure EMRs have various interesting features like practice management, electronic billing, scheduling, internationalization, patient demographics, medical reports, and patient access portal. These features enable people to adopt EMRs that results in cost reduction, capacity, and quality enhancement, and especially to lessens the disparities between wealthy poor. Therefore various open source task forces, scientists, organizations and research groups are offering open source platforms for the compilation of EMRs. Examples include OpenEMR\footnote{\url{http://www.open-emr.org/}}, OpenMRS\footnote{\url{http://openmrs.org/about/mission/}}, FreeMED\footnote{\url{http://freemedsoftware.org/}}, OSCAREMR\footnote{\url{https://oscar-emr.com/}}, OSCARMcMaster\footnote{\url{http://oscarmcmaster.16.x6.nabble.com/}} and VistA\footnote{\url{http://www.worldvista.org/}}. 

\subsubsection{Point of Care (POC) Testing}

Point of care (POC) testing or bedside testing refers to medical diagnostic testing near to the patient in the clinic or in patient's home instead of at central laboratory. It can be advantageous in emergency situations to avoid delays of sending off patients away from the point-of-care. %Generally, POC tests require less volume of samples as compared to tests conducted at the central laboratory.
%POC testing usually performed by physicians, paramedic, nurses, respiratory therapists, perfusionists, midwives and patients (glucose tests and pregnancy tests) \cite{shaw2016practical}.
POC testing devices communicate (uni- or bi-directionally) with the POC testing data management system where it is used for the decision-making process. There is a myriad of truly portable devices that are being used for POC testing by healthcare practitioner as well as by patients. For example, dipsticks are being used for the semi-quantitative estimate of a range of clinically useful analytes in urine and whole blood \cite{pugia1997comparison}.  %Samsung also made POC testing instrument called LABGEO IB10 that can measure cardiac markers \cite{scotlandevaluation}.
Similarly, another new entrant for POC teasing called PIMA\textcircled{R} is being used for the measurement of T-helper (also known as CD4 counts). Measurement of these cells is very crucial for antiretroviral HIV therapy and monitoring of immunosuppression \cite{herbert2012evaluation}. A descriptive list of various existing POC testing devices can be found in \cite{st2014existing}.

%Health 4.0: How Virtualization and Big Data are Revolutionizing Healthcare \cite{thuemmler2017case}

%\begin{itemize}
%\item Point of Care (POC) Testing
%\item Open Source EHR/EMR Systems
%\item Wearable Sensors and Implantable Nano Sensors
%\end{itemize}

\section{5G and Healthcare Opportunities}
\label{5G}
%The art of medicine regarding therapeutic interventions was advanced during nineteenth and twentieth centuries. But in these days we are able to deeply look into micro-cosmos, analyze molecular structures (DNA), and even also have stronger abilities to see at atomic and sub-atomic level. %Now, minute changes are recognizable using digital medical imaging---an ultimate diagnostic tool that makes diagnosis and treatments of diseases faster.  

\begin{figure}[!ht]
\centering
\captionsetup{justification=centering}
\centerline{\includegraphics[width=.3\textwidth]{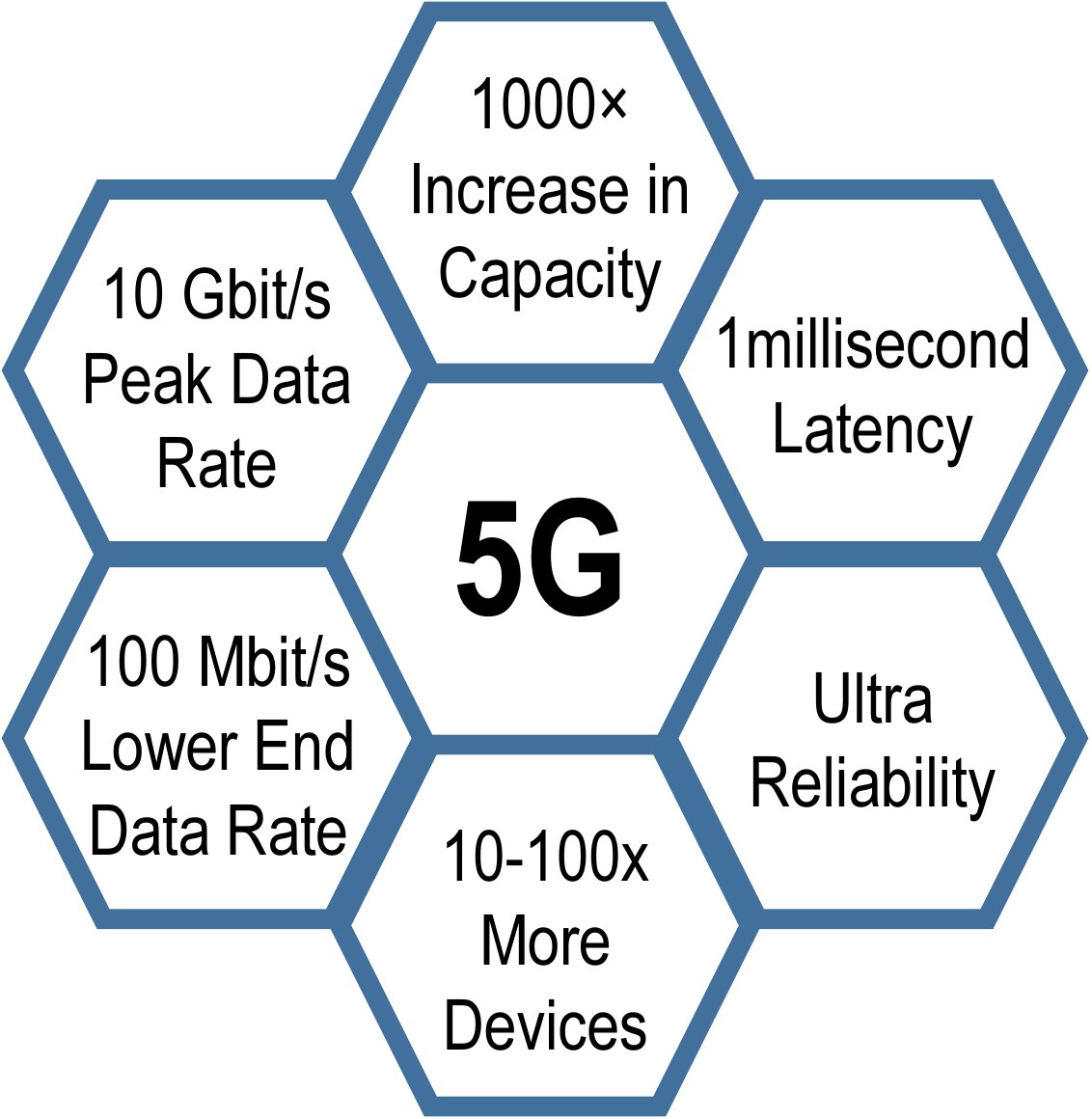}}
\caption{Some major goals articulated for 5G}
\label{fig:5gg}
\end{figure}

5G is not just the new generation of 3G, 4G technologies, but it will create a new era of agile and turbo-charge connectivity with great performance and applications tailored precisely according to the different needs of users and the economy. The goals envisioned for 5G are shown in Figure \ref{fig:5gg}. The new paradigm of 5G will also bring various innovative health services with URLLC. As a result, patient care will be improved with low physician workloads and reduced costs.  The ``Tactile Internet'' \cite{aijaz2017realizing} will enhance current transmission capabilities of touch and skills. This ultra-responsive network will enable remote physical experiences for the surgeon to provide an accurate medical diagnosis and surgery remotely. The extended transmission of multi-sensorial data including robotic touch  (i.e., haptic feedback) will improve the overall experience of real-time remote interaction and consultation. Further, virtual and augmented reality will provide an immersive user experience and can be used for virtual reality exposure therapy. These distinctive features of 5G are driving force in the future of medicine. This will change the trends in healthcare from reactive care to proactive care by providing the following missing pieces in the previous mobile communication network. 

5G will integrate various technologies in addition to those already employed by  Long-Term Evolution (LTE), and will also utilize higher-frequency bands to provide enhanced mobile broadband to address human-centric use cases. Some novel applications that will be enabled by 5G include ultra-reliable-low latency communications and massive machine-type communications for a huge number of connected devices \cite{andrews2014will}. In 5G architecture, new features---such as OFDMA (orthogonal frequency division multiple access), MIMO (multiple inputs multiple outputs), carrier aggregation, Filtered OFDM (F-OFDM), Multi-User Multiple Input Multiple Output (MU MIMO) and Sparse Code Multiple Access (SCMA)---will allow to utilize existing spectrum more effectively and increase user throughput rates and coverage. Self-organizing networks (SON) will likely to play a key role in the radio access portion. The 5G architecture will also incorporate network function virtualization (NFV), network slicing and software defined networking (SDN) to create intelligence automation that will enable quick networks scalability. 

By using these and other innovations, 5G will pave the way for a fully connected and mobile society with various promising healthcare applications (as already being conceived by various industry players \cite{sardavision} and the standardization organization 5GPP \cite{g20155g}). In what follows, we highlight some open research issues and future research directions for 5G-enabled healthcare.

\subsection{Mobile devices and tablets can help leverage AI, big data, and connectivity}

\begin{figure}[ht]
 \centering
  \captionsetup{justification=centering}
  \centerline{\includegraphics[width=.5\textwidth]{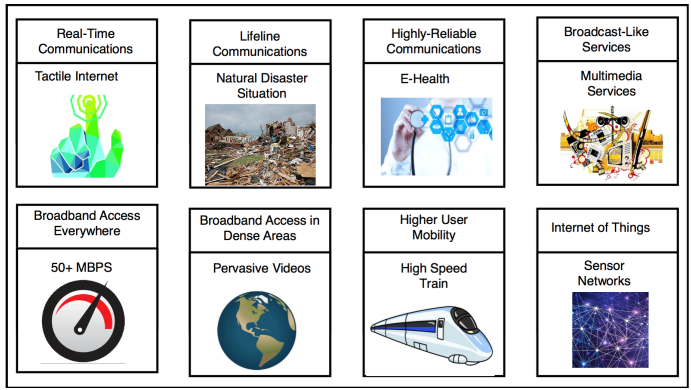}}
  \caption{5G use cases and related examples \cite{alliance20145g}}
  \label{fig:use}
\end{figure}

The proliferation of mobile communication devices and associated wireless technologies has stimulated various innovative health interventions for personalized care. The advanced connectivity of mobile devices (like mobile phones, tablets, laptops, EHRs, sensors, and personal digital assistant (PDA), etc.) and applications provide enough health related information that can excellently be utilized by big data and AI to provide smart healthcare solutions \cite{latif2017mobile}. In addition, smartphone and tablets can also serve as IoT controllers to monitor health data in real-time that can make necessary changes to physical sensors or components in medical IoT network. When 5G becomes available with improved connectivity and cloud based storage, a much wider mesh of devices, even smallest sensors, will be able to communicate. Powerful cloud based computations will bring up a mobile IoT that will provide a wide variety of opportunities, especially in healthcare. Some of the use cases along with related examples of 5G are shown in Figure \ref{fig:use}.

The world's persisting problems like uneven distribution of healthcare resources, healthcare disparities, growing number of patients with chronic diseases and increasing medical expenses will be greatly mitigated with the blend of technologies like 5G, IoT, big data and cellular technology.

%Mobile Health in the Developing World: Review of Literature and Lessons from A Case Study \cite{latif2017mobile}

%- Democratization of computing through cloud computing
%- Democratization of AI and cognitive computing (e.g., IBM Watson)
%- Commoditization of data through crowdsourcing and digital exhaust
%- High throughput and low latency connectivity. 

\subsection{5G and Universal Coverage}

Currently, 5G is in a formative stage and is anticipated to materialize by 2020. 5G will be the first technology to be developed in an age when the vision of global Internet connectivity is close to reality (as almost $60$\% of the world's population is already covered by a mobile 2G/3G/4G signal). 5G should materialize on this legacy and aim to provide universal coverage to digitize societies. By $2020$, it is expected that 5G will provide connectivity to $50$ billion devices, $212$ billion sensors and enable access to 44 zettabytes of data \cite{west20165g}. Similarly, 5G cloud is likely to cover one-third of world's population with 1.1 billion connections by $2025$\footnote{\url{https://www.gsma.com/futurenetworks/technology/understanding-5g/}}. In the context of universal access, 5G is exploring the integration of various technologies such as device-to-device (D2D) communications, IoT,  massive multiple-input multiple-output (MIMO), millimeter wave communications, and full-duplex transmissions to increase coverage in all urban and rural areas, as well as on roads and on railway tracks. Additionally, unmanned areal vehicles (UAV) and drone communications will also be utilized to provide coverage in rural areas.

%In this paradigm of 5G, devices will be able to communicate without human intervention that will facilitate more convenient, safer, and smarter living. 5G is awaited as the boundless connectivity required for all and intelligent automation and systems.% and it will deliver an edge rate of $100$ Mbps; $1000$ times enhanced capacity; high peak data rate in the range of tens of Gbps; $100$ times increased speed; round trip time of 1 ms; and 100 times increase in energy efficiency \cite{onireti2016will}. 

Despite its great potential for universal access, we have noted in our previous work \cite{onireti2016will} that many 5G technologies---including the \textit{big three 5G technologies} \cite{andrews2014will} of ultra-densificiation, millimeter wave (mmWave), and massive MIMO---are primed towards increased performance rather than coverage. In fact, technologies such as ultra-densification and mmWave will make universal coverage even more challenging; even though massive MIMO can be used for coverage gains and energy reduction if the associated computational cost of massive MIMO can be overcome. The problem of universal reliable coverage for 5G will be an important concern for healthcare applications due to the criticality of its subject domain that directly affects human lives. In particular, research on extending coverage, and providing reliability guarantees, is key for enabling the various e-health applications. In terms of technologies, energy optimization (e.g., via advanced MIMO and beamforming) will become important and will require further innovations.

\subsection{In-Home Health Monitoring}

Wearable sensors and implantable medical devices (to monitor and transmit health recorded data) will feature prominently in the future of wireless healthcare. There are already many examples of wearable/implantable medical devices (e.g. Cochlear implants, cardiac defibrillators/pacemakers, insulin pumps) that are having a large impact on patients and bringing them much ease. In the future, interfacing these wireless sensing devices with 5G will present unprecedented opportunities as well as formidable challenges. 

Currently there are several technological constraints  (i.e., low data rates, limited connectivity, and security concerns, etc.) that prevent the large-scale adoption of in-home health monitoring services. The future 5G infrastructure will open up a new set of possibilities to tackle these constraints by providing extended security along with higher bandwidth, transmission reliability, and ubiquitous access. This will provide always-available communication services for health monitoring, with a great increase in transmitting capability between health service providers and patients. 

Already, there has been exciting advances in terms of the ability to conveniently performing in-home health monitoring. For example, patients can now measure their vitals (e.g., AliveCor ECG application\footnote{\url{https://www.alivecor.com/}}) at home using mobile devices at a fraction of cost, and with great convenience, of previous options that typically entailed a visit to the local hospital. The advances in 5G can spur on a decentralization trend in which healthcare facilities can be delievered locally in homes, nursing centers, surgery centers, clinics, rehabilitation places, remote areas, and even in ambulances (e.g., when shifting critical patients to central healthcare facilities). 5G will also open up new opportunities for radar technologies to monitor the health of elderly people (i.e., fall detection and quantitative gait measurement) in a non-invasive manner both in home and clinical environments. 

\subsection{Virtual Reality + Haptic/Tactile Internet}

The Tactile Internet is an ultra-responsive and ultra-reliable network connectivity that is envisioned to transmit touch and actuation in real-time. It will revolutionize almost every segment of the society with unprecedented applications and also truly shift the paradigm from content-delivery to control communications (skill-set or labor-delivery) also called haptic communications. %The Tactile Internet will interconnect traditional wired internet, mobile internet, and IoT.
It will create an Internet of entirely a new dimension to machine-machine and human-machine interaction by providing a low latency and highly reliable, secure with a supper coverage network---that are the daunting requirements for real-time interactive systems \cite{aijaz2017realizing}. Tactile Internet underpinned by the zero-delay network will provide a virtual reality based headsets, which can facilitate doctors in performing operations through telepresence.

%The future will be defined by artificial intelligence, big data analytics, autonomous IoT, augmented or virtual reality, machine learning, supported by Tactile Internet. 
%This will change the shape of almost every field of society such as industrial automation, real-time gaming, transportation system and especially healthcare systems.

\subsection{Internet of Medical Skills}
 The exhilarating breakthrough in technology focuses on connecting societies and professionals with a great responsiveness and reaching. This will create an Internet of medical skills to transfer or share your experience and expertise over a long distance using robotics and haptic feedback. 5G is not just only the spread of connectivity, but it will also enhance the opportunity for remote training with visual and tactile communication. The doctors will become teacher and students in remote areas can follow and experience surgical procedures. Robots will be controlled by the use of haptic glove and they will transfer tactile data back to operating doctor over the distance. This exciting glimpse of future is only possible with the ultra-low latency under ten milliseconds.

%5G-enabled tactile internet \cite{simsek20165g}

%Refactoring 5G architecture to reduce end to end latency.

\section{Technology for Healthcare: Challenges/ Pitfalls}
\label{tech}
Currently, healthcare industries are mainly concerned with the increased burden of chronic disease and life expectancy, shortages of resources, regulatory requirements problems, rising patient expectations and required costs.  In the upcoming years, IoT-automation, big data analytics, cloud computing and robotics will have a great transformation in healthcare worldwide. A lot of challenges will arise from these advancements such as patients' privacy, cyber security, and data governance issues. Similarly, 5G-enabled integrated, intelligent, and massive healthcare system in every city, town, and community of the world will require enormous amounts of money, time and human capital. It is critical to adopt intelligent, informed and pre-planned strategies to avoid waste of money, time, and labor.

% Big data analytics and cloud computing will create more efficient, reliable, cost effective healthcare systems, with innovative tools to analyze an unprecedented amount of health data. 
 %In addition, healthcare organizations and vendors mostly design prototype and develop software solutions for their own purposes. Therefore, cognitive systems and workflows are required to enable better interoperability between different platforms. %Building interoperable systems involve an integration of various disparate systems, stakeholders and private and public health organization by adopting common data formats and standards. % Medical data flow through such environment will provide a 360-degreeview of health information and open up a wide variety of opportunities in healthcare. 
 
%healthcare systems will support global healthcare drive for a consistent, ubiquitous, sustainable, cost-effective and high quality of healthcare delivery. However, building an

%Healthcare is a hard nut to crack: prominent failures of Microsoft HealthVault and Google Health.

\subsection{Will computers replace doctors?}
Today's diagnoses and treatments are partially performed by patients' medical histories and symptoms. %Health practices are mostly based on averaging of the population and trial-and-result basis. 
%Different doctors propose different diagnoses and treatments for a single disease. 
A study \cite{winters2012diagnostic} found that each year $40500$ deaths are caused by misdiagnosis in the USA. Similarly, the system-related factors, e.g., poor processes, lack of teamwork, and inadequate communication, incorporated $65$\% of diagnostic error \cite{dimitrov2014systems}. These kind of diagnostic errors are adding a rise in healthcare expenditures \cite{singh2010reducing}. Most of the physicians perform check-ups and testing to suggest prescription and behavior modification that can be done better by data-driven analytics. Computers can perform better in much of diagnosis and treatments by utilizing more complex physiological and sensor data than a practitioner could comprehend. It has even been argued that computers will replace $80$\% of what doctors in the future while providing quicker, more accurate, and fact-based clinical performance \cite{khosla2012technology}. 

Notwithstanding the hype, it is safe to assume that computers will play a much larger complementary role in clinical settings in the future through the automation of the various routine matters that doctors must do, thereby creating more space for doctors to engage in non-routine health decisions that humans can manage better than algorithms. Human doctors will likely warm up to the use of information technology in the long run due to the benefits it affords. Human doctors can use their intuition that is trained over years of experience to oversee the working of algorithms and AI. Furthermore, human doctors are better equipped to manage personalized care than a machine. 

The transition to automation in health will not be easy due to the sensitive nature of the domain and the high cost associated with any kinds of errors. Furthermore, there are numerous technical challenges associated with the collection and processing of enormous amounts of data to comprehend the patients' problems and thereafter perform diagnostics through sophisticated AI and ML algorithms.

%Technology will replace 80\% of what doctors do \cite{khosla2012technology}
  
\subsection{Technological revolution needed or Behavioral revolution?}

Increasingly, the most critical challenges in healthcare are not related to information but behavioral challenges. For example, in the USA more than $36.5$\%\footnote{\url{https://www.cdc.gov/obesity/data/adult.html}} of adults have obesity. People have knowledge (eat less, exercise more) about losing weight. But behavioral changes include changing ingrained lifestyle habits such as to overcome cravings, find time and motivation to do exercises. To change the traditional healthcare trends and practices behavioral revolution is also very crucial to encourage the patients. Personal behavior has deep influences on one's health. For example, many people can make better their health by managing their chronic disease. They adopt specific health behaviors, i.e., by managing their diet plans or by regularly engaging themselves in health promotion activities \cite{ryan2009integrated}. 
Significant progress has already been made in understanding and changing health behavior \cite{ryan2008facilitating,nutbeam2010theory}, but additional technological models based on human psychology and innovative behavioral economics theories are needed---that can effectively utilize advanced technologies to meet the goals for behavioral changes among individuals and society.

%The behavioral economics of health and health care \cite{rice2013behavioral}

\subsection{Bias in Humans and Data: The Perennial Bugbear}

Every field of science experiences human biases, but in medicine, human bias is more acute and cause diagnostic inaccuracies and medical errors. Doctors mostly encounter with the common diseases, and occasionally with some distinct diseases. In many cases, different patients having the same disease react differently to medicine, and physicians' attempts to make sense and perform diagnosis from such uncertain and imprecise information may lead to wrong therapeutic decisions. Similarly, physicians apply new medical product or treatment to their patients since most of the evidence regarding such innovative medical treatments has been generated from a highly controlled research environments \cite{schneeweiss2014learning}. These new medical products and interventions can influence patients' health differently because of various disparities such as quality of life, sex, income, geographic location, education and disability status. The same biases are experienced by any predictive model or algorithm that work on data and statistical process to identify patterns. For example, machine-learning models learn from initial training data by analyzing their patterns and make predictions about the new (test) data by finding similar patterns. If these models learn wrong signals from training data, the subsequent results will lead to the wrong prediction. In \cite{mukherjee2015laws}, Mukherjee notes insightfully that the emergence of innovative medical technologies will not reduce the bias from medicine, but they will amplify it. Also, big data is not the solution of this bias, it is a source of more subtle biases. Therefore, to facilitate these biased learning, we are required such algorithms that ensure the actual effect of intervention by comparing the groups of similar patients rather than propensity score methods, which include attributes of many patients. In this way, confounding biases can be reduced from healthcare databases.

%``The advent of new medical technologies will  not  diminish  bias.  They  will  amplify  it. More human arbitration and interpretation will be needed to make sense of studies—and thus more biases will be introduced. Big data is not the solution to the bias problem; it is merely a source of more subtle (or even bigger) biases.'' Siddhartha Mukherjee \cite{mukherjee2015laws}.

\subsection{How to incentivize 5G healthcare?}
The objective of a digital healthcare system, empowered by 5G across the globe, is the benefit of consumers (patients and doctors), businesses and inclusively of the economy. 5G will pave the new ways for the provision of specialized healthcare services–-- in particular services by IoT (Telecare and Telehealth). %Such services %ensure a high level of quality, reliability, security, safety and cost appropriate for customers' expectations by 
%require a cementing strong relationships between different operators, vendors, entrepreneurs, small and medium enterprises.
Currently, healthcare providers hardly receive reimbursement for telemedicine and at home health therapies over video conferencing or mobile phone \cite{west20165g} which discourages the use of such services. This call for new policies that would promote appropriate incentives and reward (investment) for telemedicine like services and 5G orchestration platforms. These investments and incentives should be enough to operate 5G based health monitoring services and should remain productive enough to finance continuous upgradations for speed and capacity. %Therefore, Government authorities also need to update reimbursement policies for the new advances in digital medicine. In addition, generally, physicians are being incentivized by patients regardless of the quality of care they provide or whether the patients recognize it or not (fee-for-service).  
In addition, for 5G based digital healthcare system, value-based reimbursement, rather than fee-for-service, should be adopted to reward physicians by linking health outcomes with the treatment results and diagnosis records suggested by physicians. In this way, both service providers and patients will be able to observe the results of treatments that can directly correlate the worth of expenditure.

%It has been noted that physicians disproportionately change the provision of care intensify by providing elective treatments as reimbursements rise, for example, $2$\% percent increase in reimbursement rates leads to $3$\% increase in the quality of care \cite{clemens2014physicians}. 

\subsection{Security and Privacy for 5G-enabled healthcare}

5G network will link almost every aspect of human life to the Internet by using billions of devices and sensors. This will probably result in various threats to security and privacy of consumer data.  5G security threats are more serious to healthcare systems as cyber attacks on them can be detrimental on a society-level. In particular, IoT devices will be more exposed to vulnerabilities as most of the small sensors and tiny devices with low computational power unable to handle complex encryption algorithms. Consequently, the data in transit will have to be sent without any encryption. So strong mechanisms are required to secure or encrypt such bare communications. Similarly, cloud based IoT platforms used for outsourced storage and computation, due to the resource constraints of IoT sensors and devices, will also bring a series of privacy and security issues \cite{zhou2017security}. 5G networks must deal with the cybersecurity risks and privacy concerns of users, Governments, and organizations. The vision of 5G will not become reality without robust security measures that can preserve the ethical and privacy concerns.  Therefore, the level of end-to-end integrated security, confidentiality for 5G networks should be more comprehensive than previous generations of mobile networks.

\section{Conclusions} 
\label{CON}

In this paper, we have presented an overview of an impending healthcare revolution that will be empowered by 5G and concomitant technologies, such as the internet of things (IoT), big data, and artificial intelligence (AI). 5G wireless will transform the field of healthcare by augmenting  human capacity and reach in a few years by allowing resource pooling, virtualization, high-performance and reliable telemedicine, and tactile Internet with haptic feedback. 5G will enable novel healthcare applications and will also allow the ad-hoc orchestration of healthcare services by integrating patients, medical practitioners, and social workers through its support of enhanced broadband, low-latency connectivity, and ubiquitous access. We have highlighted the exciting research and implementation opportunities in building this future of 5G-enabled healthcare while also pinpointing the substantial challenges involved and the potential pitfalls.

\bibliographystyle{unsrt}
\bibliography{5G_health}

\begin{thebibliography}{10}

\bibitem{mcginnis2013best}
J~Michael McGinnis, Leigh Stuckhardt, Robert Saunders, Mark Smith, et~al.
\newblock {\em Best care at lower cost: the path to continuously learning
  health care in America}.
\newblock National Academies Press, 2013.

\bibitem{challenges}
Deloitte.
\newblock 2017 global health care outlook: Making progress against persistent
  challenges, access date: 7-{July}-2017.
\newblock
  \url{https://www2.deloitte.com/global/en/pages/life-sciences-and-healthcare/articles/global-health-care-sector-outlook.html}.

\bibitem{hayrinen2008definition}
Kristiina H{\"a}yrinen, Kaija Saranto, and Pirkko Nyk{\"a}nen.
\newblock Definition, structure, content, use and impacts of electronic health
  records: a review of the research literature.
\newblock {\em International journal of medical informatics}, 77(5):291--304,
  2008.

\bibitem{agarwal2012universal}
Dinesh Agarwal.
\newblock Universal access to health care for all: Exploring road map.
\newblock {\em Indian journal of community medicine: official publication of
  Indian Association of Preventive \& Social Medicine}, 37(2):69, 2012.

\bibitem{savedoff2009moving}
William~D Savedoff.
\newblock A moving target: universal access to healthcare services in latin
  america and the caribbean.
\newblock Technical report, Inter-American Development Bank, 2009.

\bibitem{latif2017mobile}
Siddique Latif, Rajib Rana, Junaid Qadir, Muhammad Imran, Shahzad Younis,
  et~al.
\newblock Mobile health in the developing world: Review of literature and
  lessons from a case study.
\newblock {\em IEEE Access}, 2017.

\bibitem{wu2000projection}
Shin-Yi Wu and Anthony Green.
\newblock Projection of chronic illness prevalence and cost inflation.
\newblock {\em Santa Monica, CA: RAND Health}, 18, 2000.

\bibitem{priester2005current}
Reinhard Priester, Robert~L Kane, and Annette~M Totten.
\newblock How the current system fails people with chronic illnesses.
\newblock In {\em Meeting the challenge of chronic illness}, pages 1--19. Johns
  Hopkins University Baltimore, MD, 2005.

\bibitem{world2011global}
World~Health Organization et~al.
\newblock Global health and aging.
\newblock {\em Geneva: World Health Organization}, 2011.

\bibitem{naicker2009shortage}
Saraladevi Naicker, Jacob Plange-Rhule, Roger~C Tutt, and John~B Eastwood.
\newblock Shortage of healthcare workers in developing countries--africa.
\newblock {\em Ethnicity \& disease}, 19(1):60, 2009.

\bibitem{world2014global}
World~Health Organization et~al.
\newblock Global health workforce shortage to reach 12.9 million in coming
  decades; 2013.
\newblock {\em Available on: http://www. who.
  int/mediacentre/news/releases/2013/health-workforce-shortage/en}, 2014.

\bibitem{istepanian2016m}
Robert~SH Istepanian and Bryan Woodward.
\newblock {\em M-health: Fundamentals and Applications}.
\newblock John Wiley \& Sons, 2016.

\bibitem{brook2009possible}
Robert~H Brook.
\newblock Possible outcomes of comparative effectiveness research.
\newblock {\em Jama}, 302(2):194--195, 2009.

\bibitem{dickman2016health}
Samuel~L Dickman, Steffie Woolhandler, Jacob Bor, Danny McCormick, David~H Bor,
  and David~U Himmelstein.
\newblock Health spending for low-, middle-, and high-income americans,
  1963--2012.
\newblock {\em Health Affairs}, 35(7):1189--1196, 2016.

\bibitem{wolfe2011poverty}
Barbara Wolfe.
\newblock Poverty and poor health: can health care reform narrow the rich-poor
  gap?
\newblock {\em Focus}, 28(2):12, 2011.

\bibitem{bahga2014internet}
Arshdeep Bahga and Vijay Madisetti.
\newblock {\em Internet of Things: A hands-on approach}.
\newblock VPT, 2014.

\bibitem{islam2015internet}
SM~Riazul Islam, Daehan Kwak, MD~Humaun Kabir, Mahmud Hossain, and Kyung-Sup
  Kwak.
\newblock The internet of things for health care: a comprehensive survey.
\newblock {\em IEEE Access}, 3:678--708, 2015.

\bibitem{ali2016big}
Anwaar Ali, Junaid Qadir, Raihan ur~Rasool, Arjuna Sathiaseelan, Andrej
  Zwitter, and Jon Crowcroft.
\newblock Big data for development: applications and techniques.
\newblock {\em Big Data Analytics}, 1(1):2, 2016.

\bibitem{gulamhussen2013big}
Amitte Gulamhussen, Robert Hirt, Marc Ruckebier, Jonathan Orban~de Xivry,
  Guillaume Marcerou, Jeroen Melis, et~al.
\newblock " big data in healthcare: What options are there to put the patients
  in control of their data?
\newblock In {\em EIT Foundation Annual Innovation Forum}, 2013.

\bibitem{costa2014big}
Fabricio~F Costa.
\newblock Big data in biomedicine.
\newblock {\em Drug discovery today}, 19(4):433--440, 2014.

\bibitem{lazakidou2010wireless}
Athina Lazakidou.
\newblock {\em Wireless technologies for ambient assisted living and
  healthcare: systems and applications: Systems and applications}.
\newblock IGI Global, 2010.

\bibitem{khoumbati2009handbook}
Khalil Khoumbati.
\newblock {\em Handbook of Research on Advances in Health Informatics and
  Electronic Healthcare Applications: Global Adoption and Impact of Information
  Communication Technologies: Global Adoption and Impact of Information
  Communication Technologies}.
\newblock IGI Global, 2009.

\bibitem{ng2006wireless}
HS~Ng, ML~Sim, Chor~Min Tan, and CC~Wong.
\newblock Wireless technologies for telemedicine.
\newblock {\em BT Technology Journal}, 24(2):130--137, 2006.

\bibitem{mahmoodi20175g}
Massimo Condoluci~Toktam Mahmoodi and Maria A Lema~Mischa Dohler.
\newblock 5g iot industry verticals and network requirements.
\newblock {\em Powering the Internet of Things With 5G Networks}, page 148,
  2017.

\bibitem{khosla201420}
Vinod Khosla.
\newblock 20-percent doctor included: Speculations \& musings of a technology
  optimist, 2014.

\bibitem{hafezi2015ingestible}
Hooman Hafezi, Timothy~L Robertson, Greg~D Moon, Kit-Yee Au-Yeung, Mark~J
  Zdeblick, and George~M Savage.
\newblock An ingestible sensor for measuring medication adherence.
\newblock {\em IEEE Transactions on Biomedical Engineering}, 62(1):99--109,
  2015.

\bibitem{belknap2013feasibility}
Robert Belknap, Steve Weis, Andrew Brookens, Kit~Yee Au-Yeung, Greg Moon,
  Lorenzo DiCarlo, and Randall Reves.
\newblock Feasibility of an ingestible sensor-based system for monitoring
  adherence to tuberculosis therapy.
\newblock {\em PloS one}, 8(1):e53373, 2013.

\bibitem{andreu2015wearable}
Javier Andreu-Perez, Daniel~R Leff, Henry~MD Ip, and Guang-Zhong Yang.
\newblock From wearable sensors to smart implants---toward pervasive and
  personalized healthcare.
\newblock {\em IEEE Transactions on Biomedical Engineering}, 62(12):2750--2762,
  2015.

\bibitem{taylor2008medical}
Russell~H Taylor, Arianna Menciassi, Gabor Fichtinger, and Paolo Dario.
\newblock Medical robotics and computer-integrated surgery.
\newblock In {\em Springer handbook of robotics}, pages 1199--1222. Springer,
  2008.

\bibitem{liu2014robot}
Hongqian Liu, Theresa~A Lawrie, DongHao Lu, Huan Song, Lei Wang, and Gang Shi.
\newblock Robot-assisted surgery in gynaecology.
\newblock {\em The Cochrane Library}, 2014.

\bibitem{ding2014giving}
Jienan Ding, Yi-Je Lim, Mario Solano, Kevin Shadle, Chris Park, Chris Lin, and
  John Hu.
\newblock Giving patients a lift-the robotic nursing assistant (rona).
\newblock In {\em Technologies for Practical Robot Applications (TePRA), 2014
  IEEE International Conference on}, pages 1--5. IEEE, 2014.

\bibitem{pugia1997comparison}
Michael~J Pugia, John~A Lott, Larry~W Clark, Donald~R Parker, Jane~F Wallace,
  and Todd~W Willis.
\newblock Comparison of urine dipsticks with quantitative methods for
  microalbuminuria.
\newblock {\em European journal of clinical chemistry and clinical
  biochemistry}, 35(9):693--700, 1997.

\bibitem{herbert2012evaluation}
Sophie Herbert, Simon Edwards, Gina Carrick, Andrew Copas, Christopher
  Sandford, Marc Amphlett, and Paul Benn.
\newblock Evaluation of pima point-of-care cd4 testing in a large uk hiv
  service.
\newblock {\em Sex Transm Infect}, 88(6):413--417, 2012.

\bibitem{st2014existing}
Andrew St~John and Christopher~P Price.
\newblock Existing and emerging technologies for point-of-care testing.
\newblock {\em The Clinical Biochemist Reviews}, 35(3):155, 2014.

\bibitem{aijaz2017realizing}
Adnan Aijaz, Mischa Dohler, A~Hamid Aghvami, Vasilis Friderikos, and Magnus
  Frodigh.
\newblock Realizing the tactile internet: Haptic communications over next
  generation 5g cellular networks.
\newblock {\em IEEE Wireless Communications}, 24(2):82--89, 2017.

\bibitem{andrews2014will}
Jeffrey~G Andrews, Stefano Buzzi, Wan Choi, Stephen~V Hanly, Angel Lozano,
  Anthony~CK Soong, and Jianzhong~Charlie Zhang.
\newblock What will 5g be?
\newblock {\em IEEE Journal on selected areas in communications},
  32(6):1065--1082, 2014.

\bibitem{sardavision}
Vision from orange healthcare on 5g, access date: 7-{July}-2017.
\newblock
  \url{https://5g-ppp.eu/wp-content/uploads/2015/12/5GandHealthcareEUCNCpaper.pdf}.

\bibitem{g20155g}
Christoph~Thuemmler et~al.
\newblock 5g ppp white paper on ehealth vertical sector, 2015.

\bibitem{alliance20145g}
NGMN Alliance.
\newblock 5g white paper-executive version.
\newblock {\em White Paper, December}, 2014.

\bibitem{west20165g}
Darrell~M West.
\newblock How 5g technology enables the health internet of things.
\newblock {\em Brookings Center for Technology Innovation}, 3, 2016.

\bibitem{onireti2016will}
Oluwakayode Onireti, Junaid Qadir, Muhammad~Ali Imran, and Arjuna Sathiaseelan.
\newblock Will 5g see its blind side? evolving 5g for universal internet
  access.
\newblock In {\em Proceedings of the 2016 workshop on Global Access to the
  Internet for All}, pages 1--6. ACM, 2016.

\bibitem{winters2012diagnostic}
Bradford Winters, Jason Custer, Samuel~M Galvagno, Elizabeth Colantuoni,
  Shruti~G Kapoor, HeeWon Lee, Victoria Goode, Karen Robinson, Atul Nakhasi,
  Peter Pronovost, et~al.
\newblock Diagnostic errors in the intensive care unit: a systematic review of
  autopsy studies.
\newblock {\em BMJ Qual Saf}, 21(11):894--902, 2012.

\bibitem{dimitrov2014systems}
Dimiter~V Dimitrov.
\newblock Systems patientomics: the virtual in-silico patient.
\newblock {\em New Horizons in Translational Medicine}, 2(1):1--4, 2014.

\bibitem{singh2010reducing}
Hardeep Singh and Mark Graber.
\newblock Reducing diagnostic error through medical home-based primary care
  reform.
\newblock {\em Jama}, 304(4):463--464, 2010.

\bibitem{khosla2012technology}
Vinod Khosla.
\newblock Technology will replace 80\% of what doctors do.
\newblock {\em Fortune. com http://fortune.
  com/2012/12/04/technology-will-replace-80-of-what-doctors-do}, 2012.

\bibitem{ryan2009integrated}
Polly Ryan.
\newblock Integrated theory of health behavior change: background and
  intervention development.
\newblock {\em Clinical nurse specialist CNS}, 23(3):161, 2009.

\bibitem{ryan2008facilitating}
Richard~M Ryan, Heather Patrick, Edward~L Deci, and Geoffrey~C Williams.
\newblock Facilitating health behaviour change and its maintenance:
  Interventions based on self-determination theory.
\newblock {\em European Health Psychologist}, 10(1):2--5, 2008.

\bibitem{nutbeam2010theory}
Don Nutbeam, Elizabeth Harris, and W~Wise.
\newblock {\em Theory in a nutshell: a practical guide to health promotion
  theories}.
\newblock McGraw-Hill, 2010.

\bibitem{schneeweiss2014learning}
Sebastian Schneeweiss.
\newblock Learning from big health care data.
\newblock {\em New England Journal of Medicine}, 370(23):2161--2163, 2014.

\bibitem{mukherjee2015laws}
Siddhartha Mukherjee.
\newblock {\em The Laws of Medicine: Field Notes from an Uncertain Science}.
\newblock Simon and Schuster, 2015.

\bibitem{zhou2017security}
Jun Zhou, Zhenfu Cao, Xiaolei Dong, and Athanasios~V Vasilakos.
\newblock Security and privacy for cloud-based iot: challenges.
\newblock {\em IEEE Communications Magazine}, 55(1):26--33, 2017.

\end{thebibliography}

\end{document}